\documentclass[twocolumn,pre,final]{revtex4}
\usepackage{amsmath}
\usepackage{graphicx}
\usepackage{verbatim}
\usepackage{color}
\usepackage{hyperref}
\usepackage{mathrsfs}
\begin{document}
\title{Elliptic billiard --- a non-trivial integrable system}
\author{Tao Ma, R. A. Serota}
\affiliation{Department of Physics, University of Cincinnati, Cincinnati, OH 45244-0011}
\date{March 14, 2011}

\begin{abstract}
We investigate the semiclassical energy spectrum of quantum elliptic billiard. The nearest neighbor spacing distribution, level number variance and spectral rigidity support the notion that the elliptic billiard is a generic integrable system. However, second order statistics exhibit a novel property of long-range oscillations. Classical simulation shows that all the periodic orbits except two are not isolated. In Fourier analysis of the spectrum, all the peaks correspond to periodic orbits. The two isolated periodic orbits have small contribution to the fluctuation of level density, while non-isolated periodic orbits have the main contribution. The heights of the majority of the peaks match our semiclassical theory except for type-O periodic orbits. Elliptic billiard is a nontrivial integrable system that will enrich our understanding of integrable systems.
\end{abstract}

\maketitle

\section{Level statistics of elliptic billiard}

\begin{figure}[htp]
\begin{center}
\includegraphics[width=2.1in]{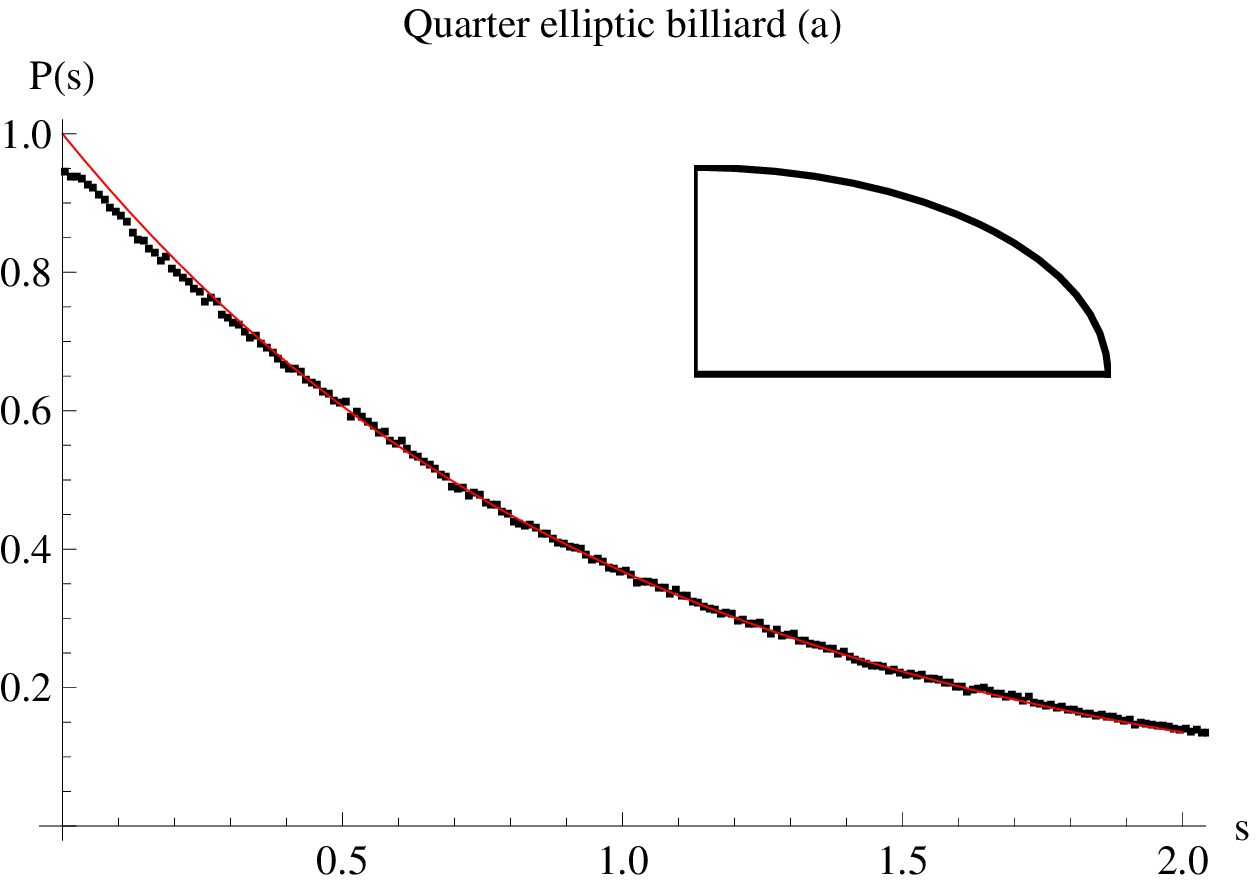}
\includegraphics[width=2.1in]{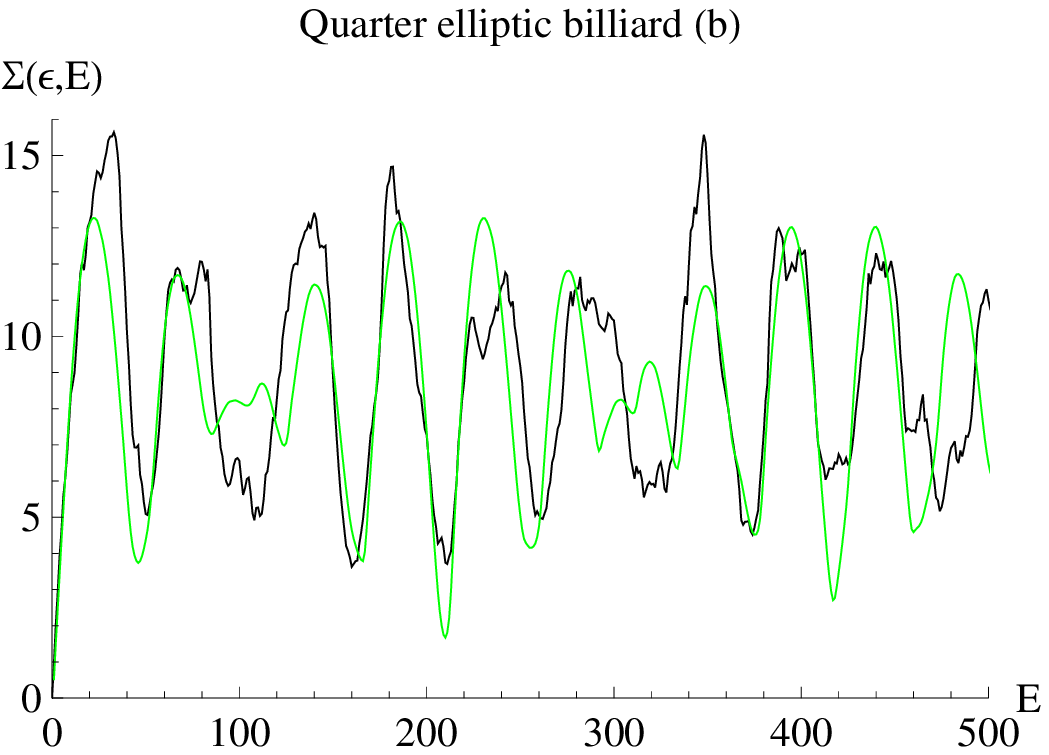}
\includegraphics[width=2.1in]{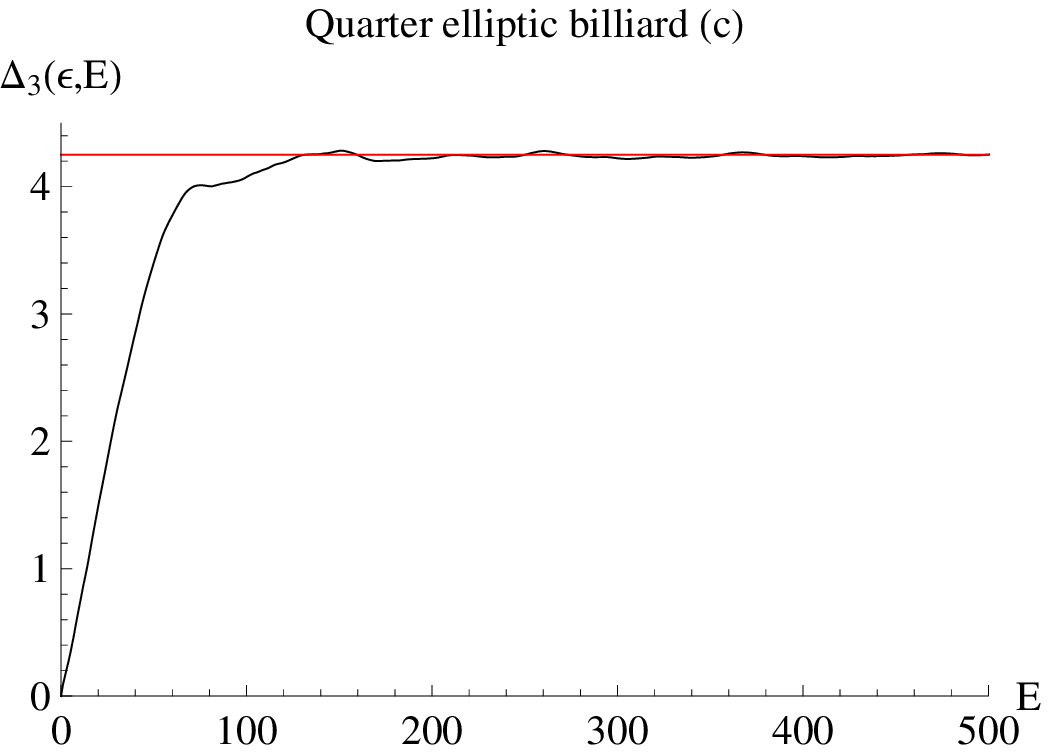}
\includegraphics[width=2.1in]{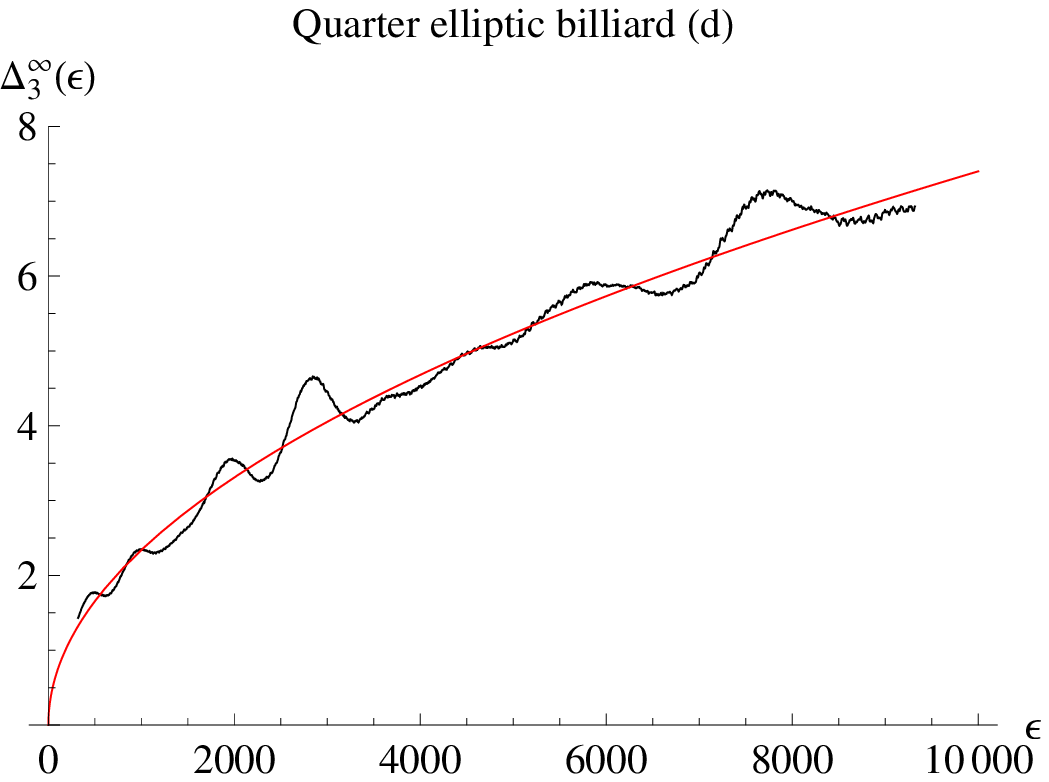}
\end{center}
\caption{\label{fig:EB} EB's $P(s)$ (a), $\Sigma$ (b), $\Delta_3(\epsilon, E)$ saturation with $E$ (c), long-range oscillations of $\Delta_3^\infty(\epsilon)$ with energy interval width (d), calculated by averaging over an ensemble of 1350 samples of $\sigma$. (a) Black dots: numerical result. Red line: Poisson distribution. Inset: shape of quarter EB. (b) Black line: numerical result. Green line: theoretical result calculated by averaging over variance calculated from 200 shortest POs. (c) Black line: numerical result. Red line: saturated value of $\Delta_3(\epsilon,E)$. (d) Black line: saturated $\Delta_3^\infty(\epsilon)$. Red line: the scaling relation in Eq. \ref{eq:scaling_relation}.}
\end{figure}

The boundary of elliptic billiard (EB) is defined as $x^2/a^2 + y^2/b^2 = 1$ with aspect ratio $\sigma = b/a$ and $ab = 1$ \cite{ayant87.I, ayant87.II, nakamura88, waalkens97}. There are four symmetry classes of eigenvalues and eigenfunctions according to the reflection symmetry of eigenfunctions against the $x$ and $y$ axes. We study the level statistics of the odd-odd class, which implies a quarter EB with Dirichlet boundary condition. The eigenvalues are computed by direct diagonalization of the Hamiltonian \cite{ayant87.I, ayant87.II, nakamura88}.

There are two kinds of averaging used for analysis of statistics in quantum chaos: ensemble averaging, e.g., over realization of disorder, and spectral averaging \cite{taoma.level.repulsion}. For integrable systems, only spectral averaging had been done before Refs. \cite{wickramasinghe05, wickramasinghe08}, in which the system parameter is sampled from algebraic numbers around the central value. A better way of sampling is sampling from a normal distribution \cite{taoma.modified.Kepler}. In this article, for EB $\sigma$ is sampled from a normal distribution centered around $1/2$ and for circular billiard (CB) around 1.

The nearest neighbor spacing distribution $P(s)$ with the level spacing $s$, level number variance $\Sigma(\epsilon, E)$ with energy $\epsilon$ and energy interval width $E$, and spectral rigidity $\Delta_3(\epsilon, E)$ are shown in Fig. \ref{fig:EB}. $P(s)$ generally follows Poisson distribution, but displays deviation at small $s$. This is a general phenomenon showing level repulsion of integrable systems \cite{taoma.level.repulsion}.
The oscillations of $\Sigma(\epsilon, E)$ are explained by the semiclassical equation
\begin{equation}
\Sigma(\epsilon, E)
= \sum_{j} \frac{8A_{j}^2}{\hbar^{N-1}T_{j}^2}\sin^2 \left(\frac{ET_{j}}{2\hbar}\right),
\end{equation}
where $j$ counts periodic orbits (PO), $\hbar$ is the Planck constant, $A_{j}$, $T_{j}$ amplitude and period of PO respectively, $2N$ the dimension of phase space \cite{wickramasinghe08}. Like rectangular billiard (RB), the match between numerical and theoretical results is achieved only after ensemble averaging in theoretical calculation \cite{taoma.modified.Kepler}. Spectral rigidity saturates. Like RB, $\Delta_3$ of EB generally follows the squared-energy scaling relation \cite{taoma.modified.Kepler}:
\begin{equation}\label{eq:scaling_relation}
\Delta_3^\infty(\epsilon) \propto \sqrt\epsilon ,
\end{equation}
but displays long-range oscillations.
All the statistics support that EB is a generic integrable system.

\subsection{spectral rigidity and global variance}

\begin{figure}[htp]
\begin{center}
\includegraphics[width=2.5in]{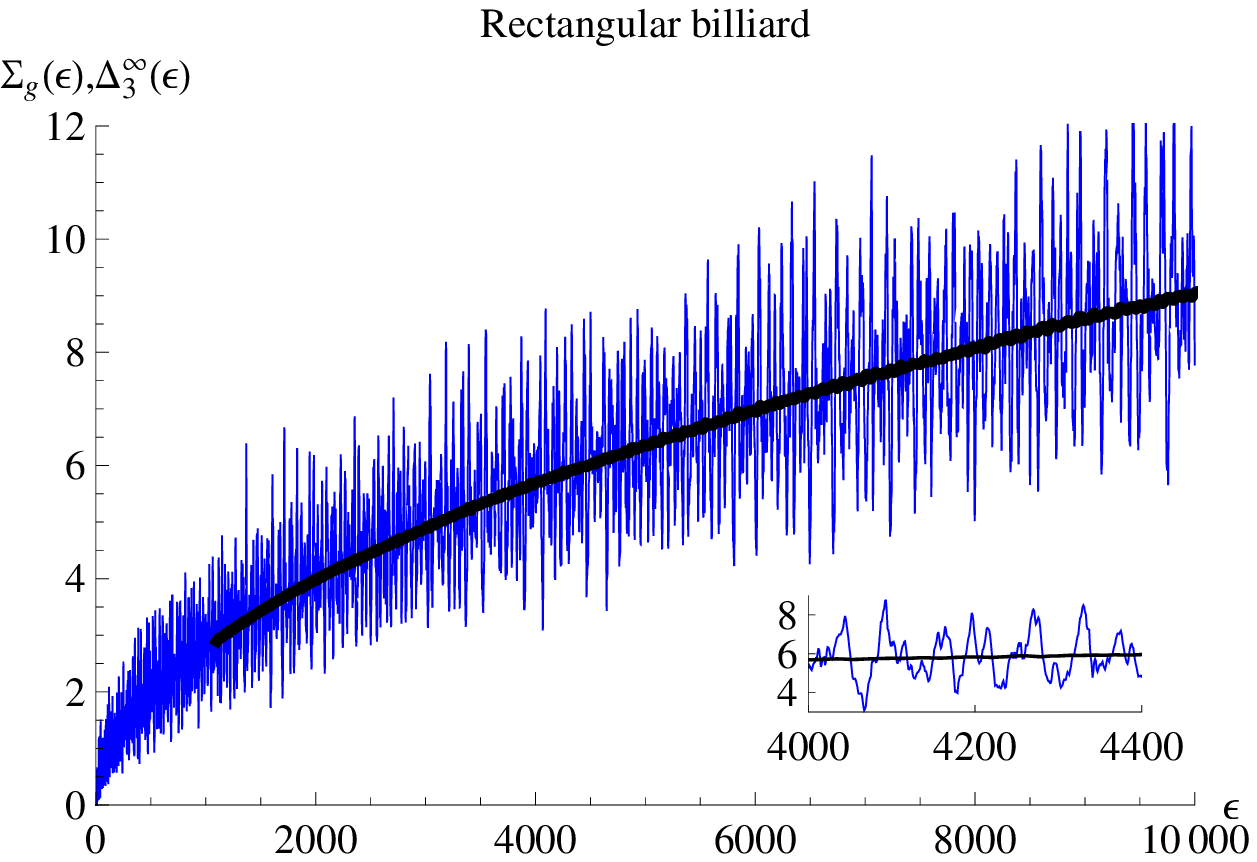} \\ \includegraphics[width=2.5in]{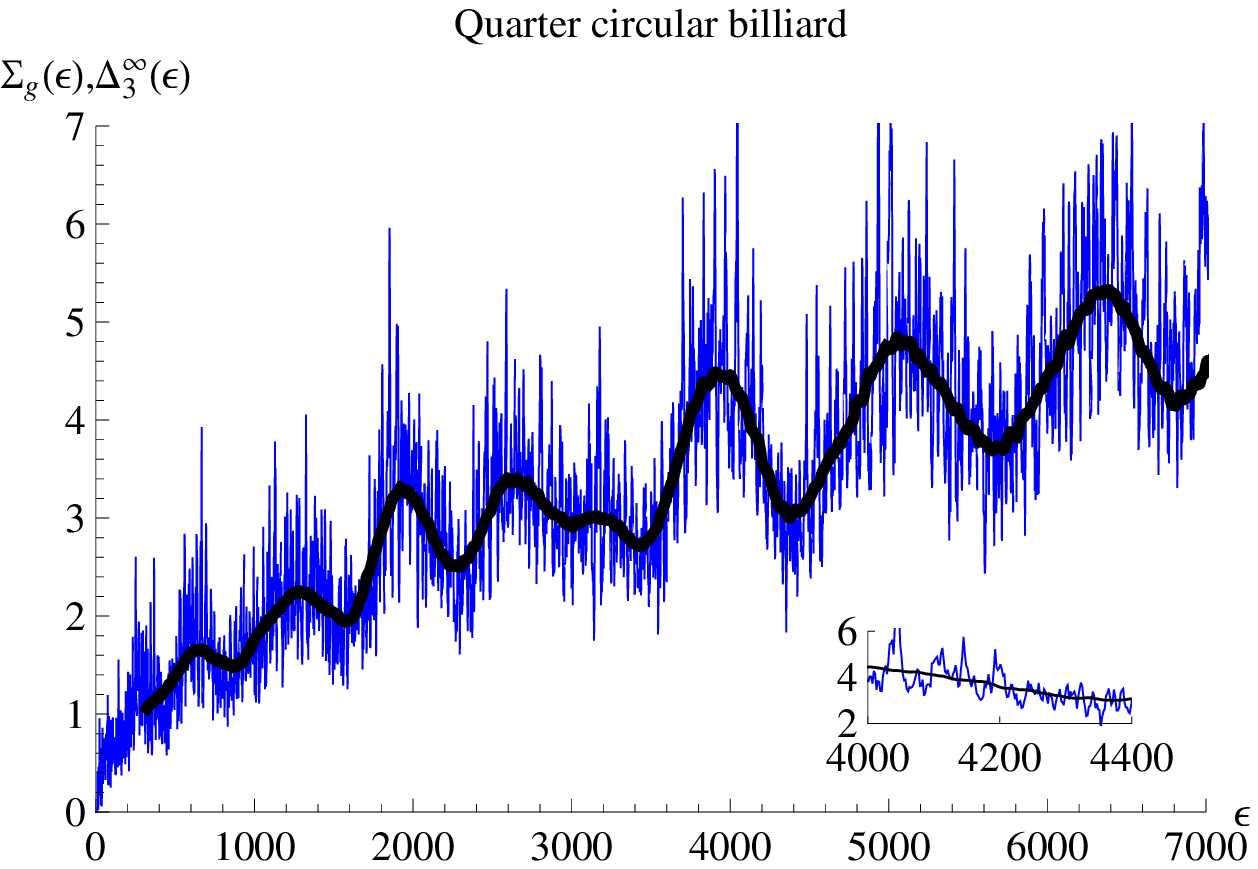} \\ \includegraphics[width=2.5in]{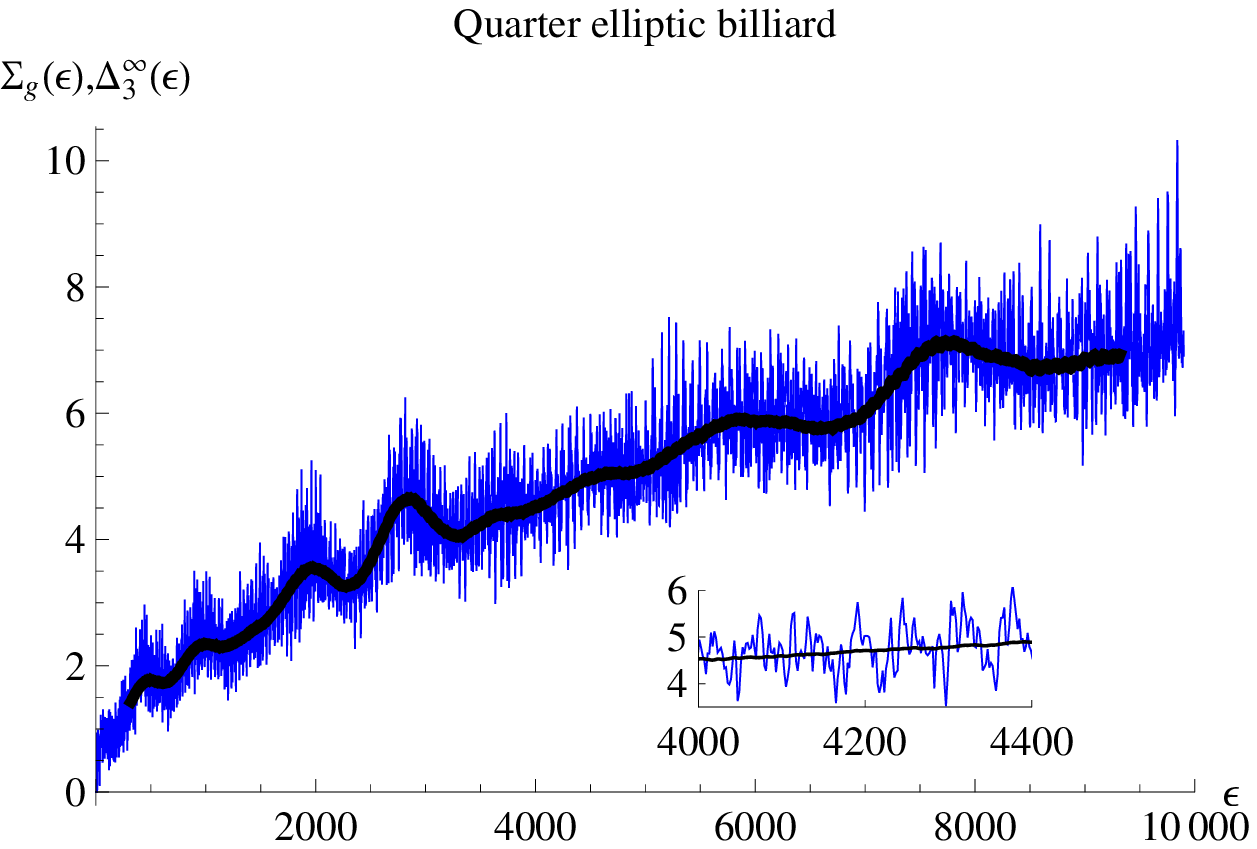}
\end{center}
\caption{\label{fig:globalVariance} Comparison between spectral rigidity and global variance. Red line: $\Delta_3^\infty(\epsilon)$. Blue line: $\Sigma_g(\epsilon)$ of RB, quarter EB, and quarter CB. The ensemble size of RB calculating $\Sigma_g$ is $10^5$. The oscillation of $\Sigma_g(\epsilon)$ is unlikely to disappear in the limit of infinite ensemble size.}
\end{figure}

The long-range oscillations of spectral rigidity show some global oscillations of level density. To reveal this global oscillations, we investigate a special case of level number variance, the global variance defined as
\begin{equation}\label{eq:global_variance}
\Sigma_g( \epsilon )
\equiv \langle [\delta\mathscr{N}(\epsilon) ]^2 \rangle
\equiv \langle [\mathscr{N}(\epsilon) - \langle\mathscr{N}(\epsilon)\rangle ]^2 \rangle,
\end{equation}
where $\mathscr{N}(\epsilon)$ is the spectral staircase. Global variance is also a special case of the correlation function of spectral staircase defined as $\langle \delta\mathscr{N}(\epsilon_1) \delta\mathscr{N}(\epsilon_2) \rangle$ \cite{serota.correlation.function}.
It is proved that the the rigidity and global variance are equal: \cite{serota.correlation.function, wickramasinghe08}
\begin{equation}\label{eq:global_variance_rigidity_relation}
\Sigma_g(\epsilon) = \sum_{j}\frac{2A_{j}^2}{\hbar^{N-1}T_{j}^2} = \Delta_3 .
\end{equation}
The above equation is partially correct as $\Sigma_g$ fluctuates around $\Delta_3$ for RB, CB and EB as demonstrated in Fig. \ref{fig:globalVariance}. For RB, $\Sigma_g$ and $\Delta_3$ follow the squared energy scaling relation. For CB or EB, the long-range oscillations of $\Sigma_g$ and $\Delta_3$ are ``synchronized''. The scales of oscillations of $\Sigma(\epsilon,E)$ and $\Delta_3^\infty(\epsilon)$ are different. The former is $E \sim \sqrt{\epsilon}$, and the latter is longer than this.

We think the long-range oscillations of $\Delta_3$ and $\Sigma_g$ originate from collective effect of type-R POs of EB or CB and exist in any billiard with whispering gallery modes. The whispering gallery POs have approximately equal orbit length or multiples of that and hence they coherently cause global fluctuation of level density.

\section{Periodic orbits of elliptic billiard}

\begin{figure}[htp]
\begin{center}
\includegraphics[width=1.5in]{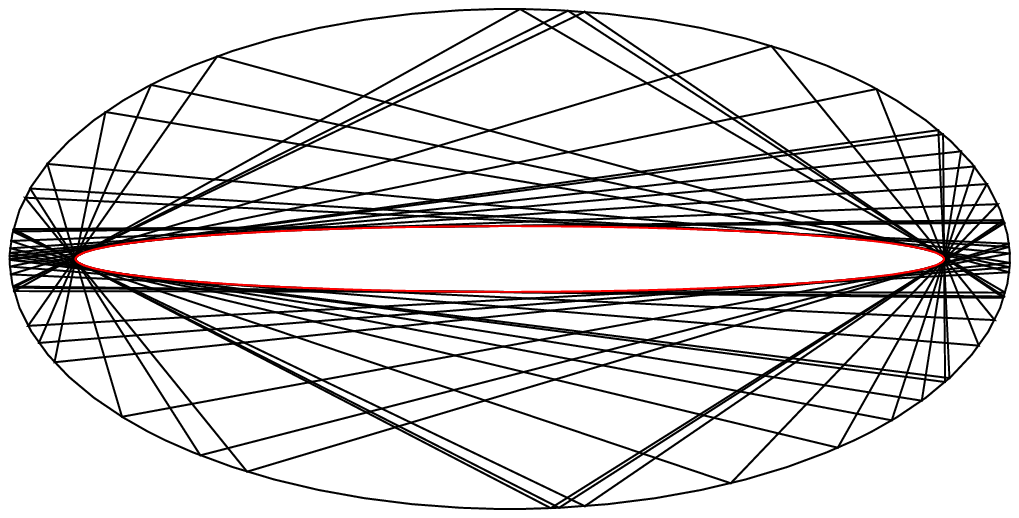}
\includegraphics[width=1.5in]{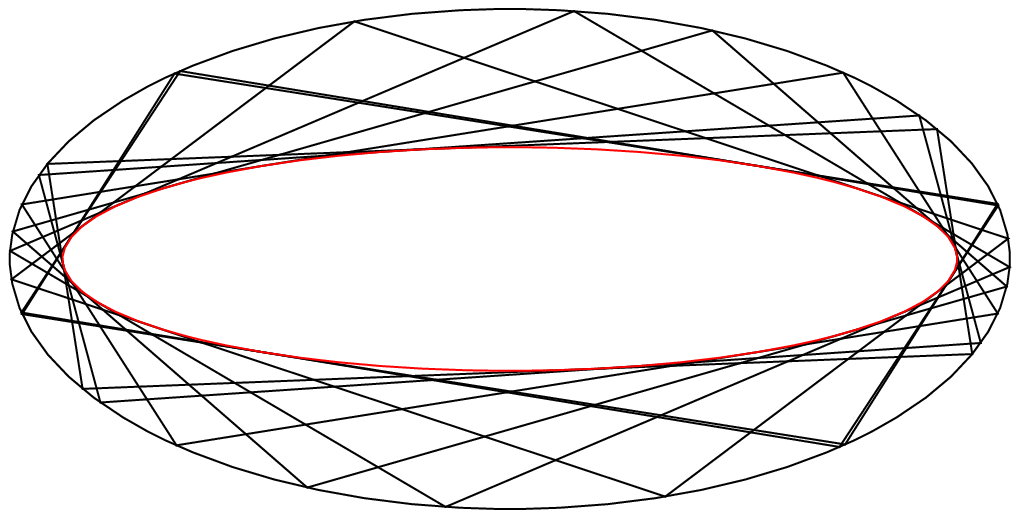}
\includegraphics[width=1.5in]{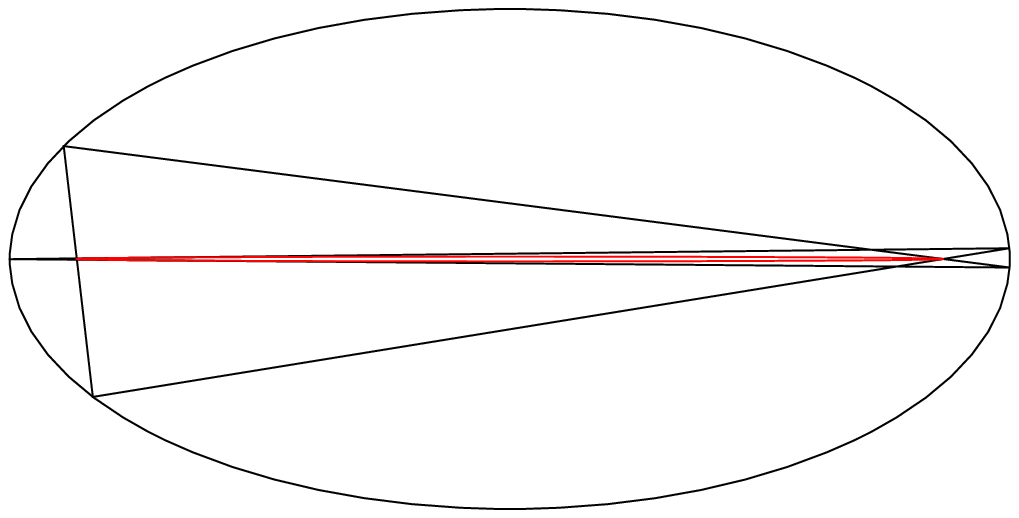}
\includegraphics[width=1.5in]{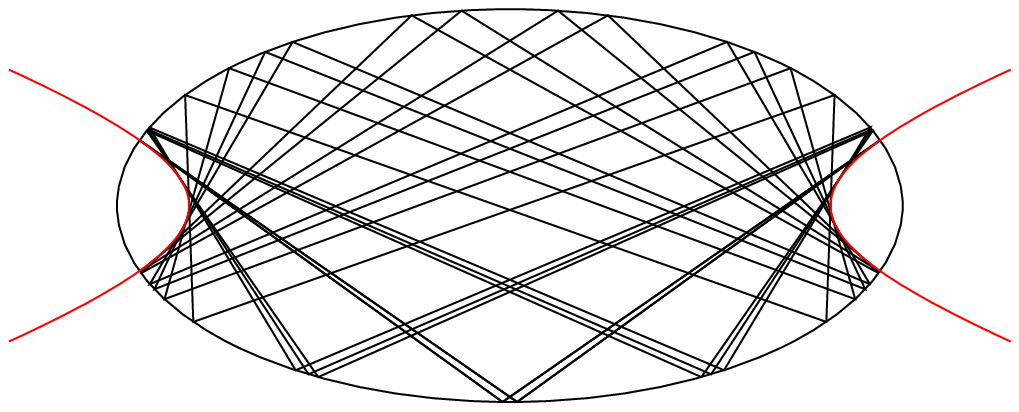}
\caption{\label{fig:orbitsR345}
$R_{3,1}$, $R_{4,1}$, $R_{5,2}$, and $O_4$ families of POs.
Red line: envelope ellipse or hyperbola. }
\end{center}
\end{figure}

Semiclassical spectral properties are closely related to classical POs.
We have the following theorems concerning classical POs of EB, demonstrated in Fig. \ref{fig:orbitsR345}.
First, the numerical work of hundreds of families of POs verifies that all the POs except two on axes are continuous. Inside the area covered by a family of POs, there are four directions to find POs in the family. This makes the factor $c$ in Eq. \ref{eq:amplitude_PO_CB} below same for every continuous families of POs.
Second, the envelope of a family of POs is an ellipse or a hyperbola.
Third, the envelope ellipses, hyperbolas and EB are confocal. In Fig. \ref{fig:orbitsR345}, the foci of the envelope ellipse for $R_{3,1}$ are at $x = \pm\sqrt{1.228268^2 - 0.09297^2} = \pm \sqrt{3/2}$, which is same as EB; the foci of the envelope hyperbola of $O_4$ are at $x = \pm\sqrt{1.1547^2 + 0.408248^2} = \pm\sqrt{3/2}$.
Fourth, all the POs of a family with the same envelope curve have the same orbit length, for example all the $R_{3,1}$ POs have orbit length 6.0322.
Fifth, for type-R POs, every orbit line is tangent with the envelope ellipse and for type-O either the orbit line or its extended line is tangent with the envelope hyperbola.

\section{Fourier analysis of spectrum of rectangular, circular billiards}

To reveal the underlying POs of semiclassical spectrum, we perform Fourier analysis on the spectrum. In this section and the section below, we do not perform ensemble averaging on the Fourier analysis. The positions of peaks of Fourier analysis are identified with orbit lengths. Numerical work of EB confirms that all the peaks have the same half-width. Hence the height of a peak is proportional to amplitude of a peak and proportional to the amplitude of a PO.
\begin{equation}\label{eq:height_peak_vs_amplitude_PO}
\textit{height of peak} \propto
\textit{amplitude of peak} \propto
\textit{amplitude of PO} .
\end{equation}

\subsection{Fourier analysis of spectrum}

The Fourier analysis of spectrum has been successfully applied to analyze Sinai billiard, hydrogen atom in a strong magnetic field \cite{stockmann00}.
The trace formula expresses the fluctuation of level density as a summation over (families of) POs:
\begin{equation}\label{eq:trace_formula}
\rho(k) = \sum_{j} A_{j}( L_{j} ) e^{ikL_{j}+\alpha},
\end{equation}
where $L_{j}$ is orbit length of a family of POs and $\alpha$ the Maslov index. We do not consider $\alpha$ in this article. If we extend $A_{j}( L_{j} )$ to $A( l )$: the amplitude of POs over orbit length by extending $L_{j}$ to the whole line of $l$, we have
\begin{equation}
\rho(k)
= \int_{-\infty}^\infty A( l )  e^{ikl} dl.
\end{equation}
The reverse of a PO has orbit length $-l$.
The inverse Fourier analysis gives the amplitude
\begin{equation}\label{eq:inverse_trace_formula_smoothed}
A( l )
= \frac{1}{2\pi} \int_{-\infty}^\infty \rho(k)  e^{-ikl} dk.
\end{equation}
The discreteness of spectrum makes Fourier analysis into a summation over eigen-momentum $k_i$, where $i$ counts eigenvalues:
\begin{equation}\label{eq:inverse_trace_formula_discrete}
A( l )
= \frac{1}{2\pi} \sum_{i} e^{-ik_i l},
\end{equation}
and
\begin{equation}
A_{j}( L_{j} )
= \frac{1}{2\pi} \sum_{i} e^{-ik_i L_{j}}.
\end{equation}
For simplicity, we ignore the subscript $j$ below.

\subsection{rectangular billiard}

\begin{figure}[htp]
\begin{center}
\includegraphics[width=3.0in]{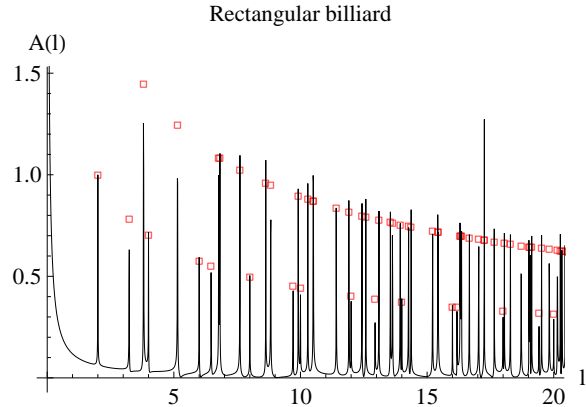}
\caption{\label{fig:Fourier_rectangle}
Black line: Fourier analysis of RB with two sides as 1 and $(\sqrt{5}+1)/2$. Red square: semiclassical theory. The $L=2$ peak is normalized to 1 in both numerical and theoretical calculations. }
\end{center}
\end{figure}

A direct quantum mechanical calculation proves that for RB the amplitude of a family of POs is given by
\begin{equation}\label{eq:amplitude_PO_RB}
A(L) \propto \frac{c}{\sqrt{L}},
\end{equation}
where the momentum space factor $c = 1/2$ for POs parallel to boundary, and 1 otherwise \cite{wickramasinghe05, taoma.modified.Kepler} as for the former there are two directions to find POs from the same family while for the latter there are four directions. Although Eq. \ref{eq:amplitude_PO_RB} is simple, it gives correct position and height of almost every peak shown in Fig. \ref{fig:Fourier_rectangle}. The rare large discrepancy for some POs is caused by interference of close-by POs.

\subsection{circular billiard}

\begin{figure}[htp]
\begin{center}
\includegraphics[width=3.0in]{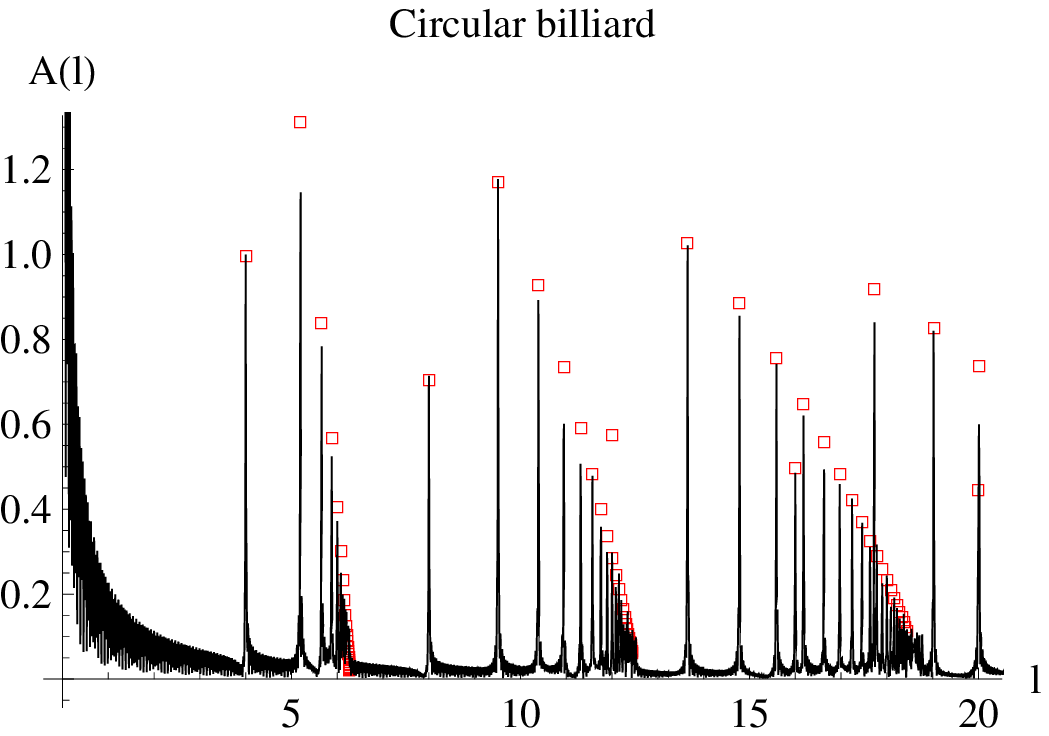}
\caption{\label{fig:Fourier_circular} Black line: Fourier analysis of CB. Red square: semiclassical theory. The $L=4$ peak is normalized to 1 in both numerical and theoretical calculations. }
\end{center}
\end{figure}

For CB, different POs cover different areas in position space. To account for this difference, we add the factor $S$, the area covered by a family of POs in position space (in the case of EB, the area covered by orbit lines in Fig. \ref{fig:orbitsR345}) and the amplitude of POs reads
\begin{equation}\label{eq:amplitude_PO_CB}
A(L) \propto \frac{c S}{\sqrt{L}},
\end{equation}
where the momentum space factor $c = 1/2$ for POs along diameters, and 1 otherwise as for the former there are two directions to find POs from the same family while for the latter there are four directions. For a family of POs denoted by $(n, m)$ with $m\leq n/2$, where in a period, the particle collides $n$ times with the boundary and makes $m$ revolutions,
\begin{eqnarray}
&S = \pi[ 1 - \cos^2(\frac{m\pi}{n})^2 ] \\
&L = 2 n \sin(\frac{m\pi}{n}) .
\end{eqnarray}
Eq. \ref{eq:amplitude_PO_CB} is verified in Fig. \ref{fig:Fourier_circular} except for interfering close-by orbits.

\subsection{summary}

The position and height of peaks of RB and CB are correctly given by the amplitude formulae of POs in Eqs. \ref{eq:amplitude_PO_RB} and \ref{eq:amplitude_PO_CB}. From our knowledge, the factor $S$ in Eq. \ref{eq:amplitude_PO_CB} has never been pointed out before. Its existence in EB is also confirmed below. An argument to justify Eqs. \ref{eq:amplitude_PO_RB} and \ref{eq:amplitude_PO_CB} is as follows.

The amplitude of a family of POs is the linear superposition of all the individual POs in the family. For integrable systems, we assume that the amplitude of each PO is only decided by its orbit length and given by $1/\sqrt{L}$, which is derived theoretically in the case of RB or more generally from Berry-Tabor formula \footnote{The factor $1/T$ in $A^2\propto 1/T$ with period $T$ in the appendix of \cite{wickramasinghe08} gives the same result.}. In this assumption, orbit stability plays no role. All the POs of a family contribute equally to the amplitude. The factor $cS$ is used to quantify how many POs are in a family. $c$ gives the number of POs in the momentum space and $S$ in the position space. For RB, every family of POs covers the whole billiard and $S$ plays no role therein.

Now we have a trace formula for integrable systems from Eq. \ref{eq:trace_formula}:
\begin{equation}\label{eq:trace_formula_generic_integrable_system}
\delta\rho(k) = \sum_{j} \frac{c S}{\sqrt{L}} e^{ikL + \alpha} .
\end{equation}
Given its similarity with Gutzwiller trace formula \cite{gutzwiller90}, this formula gives some new information compared with Berry-Tabor trace formula \cite{berry1977calculating.bound.spectrum}. It is worthwhile to understand its relation with the latter.

\section{Fourier analysis of spectrum of elliptic billiard}

\begin{figure}[htp]
\begin{center}
\includegraphics[width=3.3in]{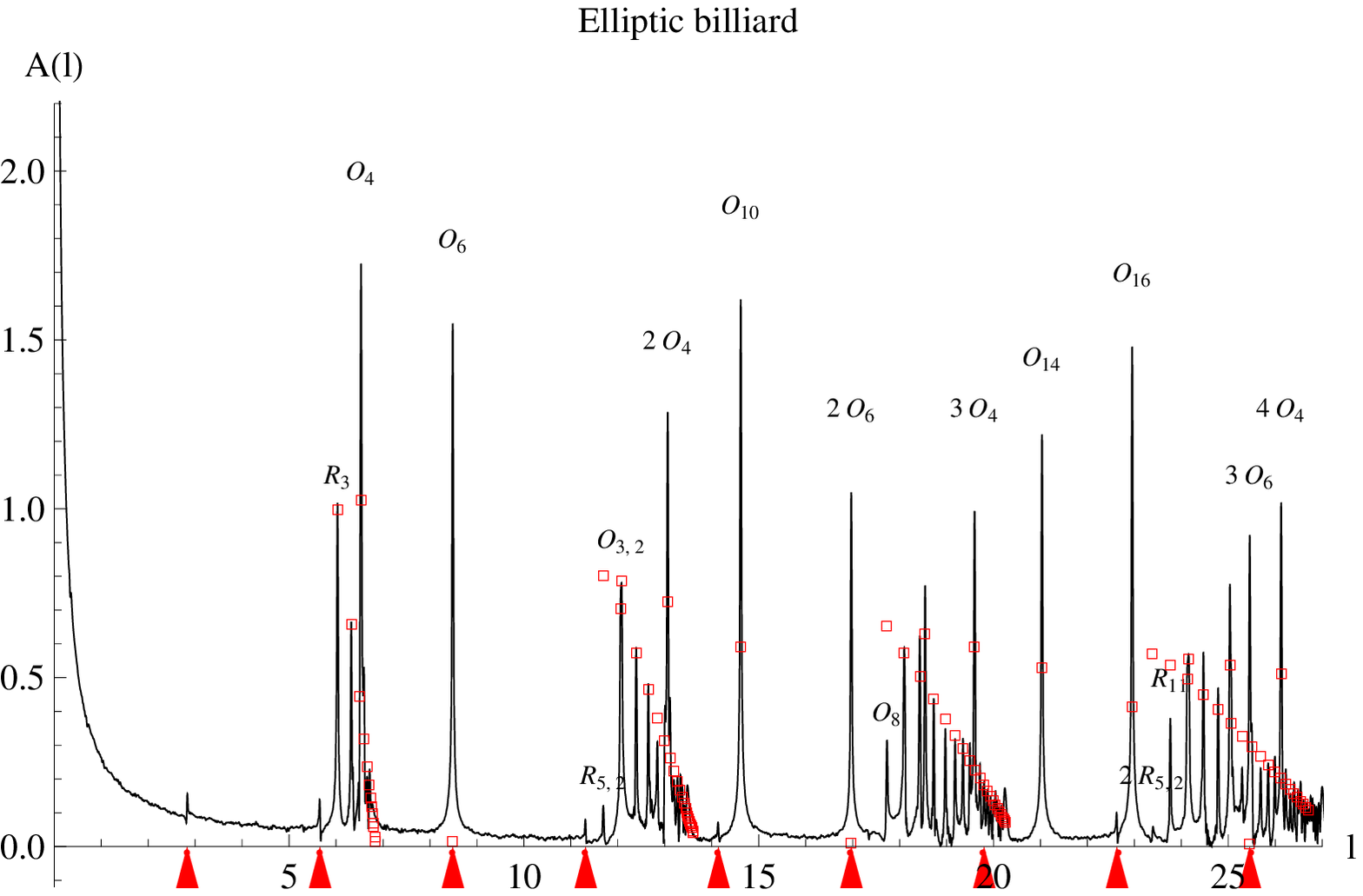}
\end{center}
\caption{\label{fig:Fourier.Theory.EB} Red arrows: positions of unstable and stable isolated POs. The Fourier analysis contains eigenvalues of all the four symmetry classes of EB with $\sigma = 1/2$ but the almost degenerate eigenvalues are removed by only considering one of them. The heights of peaks are normalized by defining the peak for $R_{3,1}$ equal to 1 in both numerical result and theory.}
\end{figure}

Comparison of Fourier analysis of EB with the semiclassical theory in Eq. \ref{eq:amplitude_PO_CB} is shown in Fig. \ref{fig:Fourier.Theory.EB}. The positions of peaks match well with very small discrepancy. At $L = 2.83$, the short peak is due to the unstable PO with $L = 4b$, indicating that the unstable isolated PO is unimportant to the level fluctuation. At $L = 5.66$, the stable isolated PO with $L = 4a$ contributes another short peak. The isolated POs contribute little to level fluctuation, which was already demonstrated in the modified Kepler problem containing one stable isolated PO \cite{taoma.modified.Kepler}. The unimportance of isolated POs can be explained by our argument in the last section as its position space factor $S = 0$. At $L = 6.32$, the continuous $R_{4,1}$-POs contribute significantly. Around $L = 6.6$, there are many peaks overlapping together. These are the contribution from $R_{n,1}$ with $n\gg 1$. These POs has the orbit length around the perimeter equal to 6.85.

For heights of peaks, good match between numerical and theoretical results is achieved for type-R POs except for $R_{5,2}$ but not for type-O POs. Two kinds of discrepancies exist. First, for $O_4$, $O_{8}$, $O_{10}$, $O_{14}$, $O_{16}$, the discrepancy is a factor around 1. Second, for $O_6$ the discrepancy is very large. For the equally spaced $O_4, 2O_4, 3O_4, 4O_4$ (repetition of $O_4$) and $O_6, 2O_6, 3O_6$ (repetition of $O_6$), their relative magnitudes are correctly given by the factor $1/\sqrt{L}$.

\section{Conclusions}

Although EB is a generic integrable system, it contains several new characteristics. The second order statistics including spectral rigidity and global variance display long-range oscillations, not noticed in other systems before. We think such oscillations exist for any billiard with whispering gallery modes. The global variance seems to be a statistical instrument worth further studying. The Fourier analysis of spectrum gives the position and amplitude of POs. For RB and CB, the numerical results match our proposed theory of amplitude of POs. For EB, only the type-R POs fully match our theory. Further work is needed on type-O POs. Future work will address the orbital magnetic response of EB. Type-R and type-O POs enclose different magnetic flux and hence they should show different orbital magnetic response. In summary, EB is a non-trivial integrable system that will further enrich our understanding of quantum chaos.

\bibliography{reference}
\end{document}